\documentclass[runningheads,a4paper]{llncs}

\usepackage[T1]{fontenc}
\usepackage[utf8]{inputenc}

\usepackage[british]{babel}
\usepackage[babel]{csquotes}
\MakeAutoQuote{“}{”}
\MakeAutoQuote*{‘}{’}

\usepackage{makeidx}  
\usepackage{times,amsmath,epsfig}
\usepackage{epstopdf}
\usepackage{amsmath}
\usepackage{lstsemantic}  
\usepackage{xcolor}
\usepackage{colortbl}
\usepackage{rotating}
\usepackage{enumerate}
\usepackage{url}
\usepackage{listings}
\usepackage{tabularx}
\usepackage{hyperref}

\usepackage{subfigure}

\usepackage{amssymb}
\usepackage{graphicx}
\usepackage{float}
\usepackage{algorithm}

\AtBeginDocument{}

\usepackage{algorithm}
\usepackage[noend]{algpseudocode}

\makeatletter
\def\BState{\State\hskip-\ALG@thistlm}
\makeatother

\usepackage{url}
\newcommand{\keywords}[1]{\par\addvspace\baselineskip
\noindent\keywordname\enspace\ignorespaces#1}

\usepackage{listings}
\usepackage{color} 
\definecolor{grey}{rgb}{0.9,0.9,0.9} 
\lstnewenvironment{lqml}[2][]
  {
    \lstset{
      basicstyle=\scriptsize \ttfamily,
      breaklines=true,
      captionpos=b,
      frame=bottomline,
      extendedchars=true,
      caption=#1,
      label=#2,
      morekeywords={def,metric,label,description,match,action,map,count,typeof,finally,unique,actionresult,ratio},
      escapechar=@,
    }
  }
   {
  }

\usepackage[disable]{todonotes}

\begin{document}

\mainmatter 

\titlerunning{Luzzu Quality Metric Language}
\title{Luzzu Quality Metric Language -- A DSL for Linked Data Quality Assessment}

\author{Jeremy Debattista, Christoph Lange, S\"{o}ren Auer}
\institute{University of Bonn \& Fraunhofer IAIS \\
\email{\{debattis,langec,auer\}@cs.uni-bonn.de}
}

\maketitle

\begin{abstract}
The steadily growing number of linked open datasets brought about a number of reservations amongst data consumers with regard to the datasets' quality.
Quality assessment requires significant effort and consideration, including the definition of data quality metrics and a process to assess datasets based on these definitions.
Luzzu is a quality assessment framework for linked data that allows domain-specific metrics to be plugged in.
In this paper we describe the Luzzu Quality Metric Language (LQML), a domain specific language (DSL) whose purpose is to enable non-programming domain experts to define quality metrics for the assessment of linked open datasets.
LQML offers notations, abstractions and expressive power, focusing on the representation of quality metrics.
It provides expressive power for defining sophisticated quality metrics.
Its integration with Luzzu enables their efficient processing and execution and thus the comprehensive assessment of extremely large datasets in a streaming way.
We also describe a novel ontology that enables the reuse, sharing and querying of such definitions.
Finally, we evaluate the proposed DSL against the cognitive dimensions of notation framework.


\end{abstract}


\section{Introduction}
\label{sec:introduction}
In May 2007 the first version of the Linked Open Data cloud~\cite{LOD-cloud} was published.
Following Linked Data principles\footnote{\url{http://www.w3.org/DesignIssues/LinkedData.html}}, 12 datasets were initially added to the LOD cloud.  
The cloud saw an increase during the years, and the data provider\footnote{Each data provider, identified by its pay-level domain (e.g. \url{http://dbpedia.org}), can have more than one dataset; 1014 datasets were discovered in 2014} count is 570 (as of August 2014).
Furthermore, an investigation (from February 2015) at the data catalogue portal datahub.io showed that 1,309 out of 9,262 datasets are tagged with the \textit{lod} or \textit{format-rdf} tags.
Moreover, other linked datasets might be available in the Web of Data but are not catalogued, thus they are not easily discovered by data consumers.
Schmachtenberg et al.~\cite{Schmachtenberg2014LodCloud} compiled a list of uncatalogued datasets from various W3C mailing list communications and the 2012 Billion Triple Challenge.

However, the increase of linked open datasets brought about a number of reservations amongst data consumers with regard to the datasets' quality.
Hitzler and Janowicz~\cite{journals/semweb/HitzlerJ13} state that linked open datasets have a reputation of being of poor quality.
Consequently, if we manage to systematically assess and improve data quality of LOD, many Linked Open Data applications can be better supported and contribute to establishing the 4th Big Data aspect -- Veracity.

Indubitably, quality assessment requires a lot of effort and consideration before processing a dataset.
Quality is commonly described as \emph{fitness for use}~\cite{Juran1974}.
Although domain experts can decide on a number of quality factors of a particular dataset, in the end it is up to data consumers to see if a dataset is suitable for their use case or not.
Providing measures on the quality of a published linked open dataset should be part of the publishing lifecycle.
Various research work defines a number of quality factors pertinent to linked open datasets~\cite{Bizer2008:PhDThesis:biblatex,Flemming2011,Hogan2012:LDC}.
Zaveri et al.~\cite{Zaveri2012:LODQ} provide an overall systematic review of such quality metrics.
Additionally, domain specific quality metrics are required to have a more comprehensive view of the dataset's quality.
For a cultural heritage dataset, for example, the ratio of resources being linked to an \emph{Integrated Authority File} (e.g. \emph{GND}\footnote{German ``Gemeinsame Normdatei''; see \url{http://www.dnb.de/EN/gnd}}) is of crucial importance.

\emph{Luzzu}\footnote{\url{http://eis-bonn.github.io/Luzzu}}, is an extensible quality assessment framework for linked open datasets.
Linked Data quality metrics can be added to Luzzu by various third parties (such as programmers and data enthusiasts).
It often occurs that \emph{data scientists}, whose spectrum ranges from data publishers and consumers to domain experts and knowledge engineers, might not be confident in programming using traditional third generation languages.
Nevertheless, they are considered to be the ideal drivers for defining domain specific quality metrics, which can be used on linked open datasets.

The main contribution of this article is the definition and implementation of the \emph{Luzzu Quality Metric Language} (LQML), a domain specific language (DSL) that enables declarative definition of quality metrics for Luzzu (cf. Section~\ref{sec:lqml}).
LQML offers notations, abstractions and expressive power, focusing on a the representation of quality metrics for Linked Dataset assessment.
A particular challenge in the definition of LQML was to balance between providing the expressive power for defining sophisticated quality metrics on the one hand and ensuring their efficient processing and execution on the other hand.
As a result, our implementation enables the comprehensive assessment of extremely large datasets with respect to many quality metrics in a streaming way.
LQML is designed in a way that metrics can be written by non-programmers that are experts in the domain (cf.\cite{vanDeursen:2000:DLA:352029.352035,dsl-little-languages,Mernik:2005:DDL:1118890.1118892}).
Hudak~\cite{dsl-little-languages} suggests that DSLs have the potential to improve productivity in the long run and with LQML we aim to contribute to overcoming one of the main problems of Linked Open Data -- data quality.

We also define a new vocabulary to enable the reuse, sharing and querying of LQML metrics in a semantic manner (cf. Section~\ref{sec:extending}).
The usability of the LQML is systematically assessed (cf. Section~\ref{sec:evaluation}) against the ``cognitive dimensions of notation'' (CD) evaluation framework.
These dimensions provide a comprehensive view of how users can manage and use a defined language.
We also briefly outline the state-of-the-art in domain specific languages (cf. Section~\ref{sec:related}) and concluding remarks and an outlook on future work are discussed in Section~\ref{sec:conclusion}.




\section{Luzzu Quality Metric Language}
\label{sec:lqml} 
The \emph{Luzzu Quality Metric Language} (LQML) is a structural declarative language that enables the definition of quality metrics (called \emph{blueprints}) in Luzzu.
Based on our experience from the use cases of the DIACHRON FP7 EU project\footnote{\url{http://diachron-fp7.eu}}, we anticipate that most domain-specific quality metrics are very similar structure-wise, with minor changes required only in the rules' conditions.

\subsection{Analysis}
\label{sec:analysis}
Data quality assessment varies from one domain to another.
Although there exist a number of generic quality metrics as defined in~\cite{Zaveri2012:LODQ}, different domains might require the assessment of different features.
For example, where in geographical datasets the properties \texttt{geo:long} and \texttt{geo:lat} are absolutely required for resources that are defined as a place (such as country and city), these properties might be redundant in health oriented datasets.
The idea of LQML is that data scientists can define various quality metrics over a dataset (or a domain of datasets).
These declarative definitions are translated into Java byte-code (see Section~\ref{sec:impl}) and integrated within the Luzzu framework.
For the proposed domain specific language, we identified a domain terminology based on quality metrics required by pilot partners in the DIACHRON project.

\paragraph{Use case overview}
\begin{description}
\item[EBI – ] One of the services of the European Bioinformatics Institute (EBI\footnote{\url{http://www.ebi.ac.uk}}) is to provide linked datasets to the scientific community, with their main development focusing around the Experimental Factor Ontology (EFO).
The EFO ontology is then used to annotate data in databases at the EBI.
EFO is an evolving ontology by nature and concepts from external ontologies are constantly being added (or replaced) in the EFO.
\item[Data Publica – ] This French company\footnote{\url{http://www.data-publica.com}} provides a number of data services, which include the management of the largest and most complete directory of electronic data in France.
This directory covers all data available in France (private and public), annotating it with relevant metadata, and making it available to the public through various means (search engines, visualisations, etc.).
\end{description}

\begin{figure}[t]
\center
\includegraphics[width=\columnwidth]{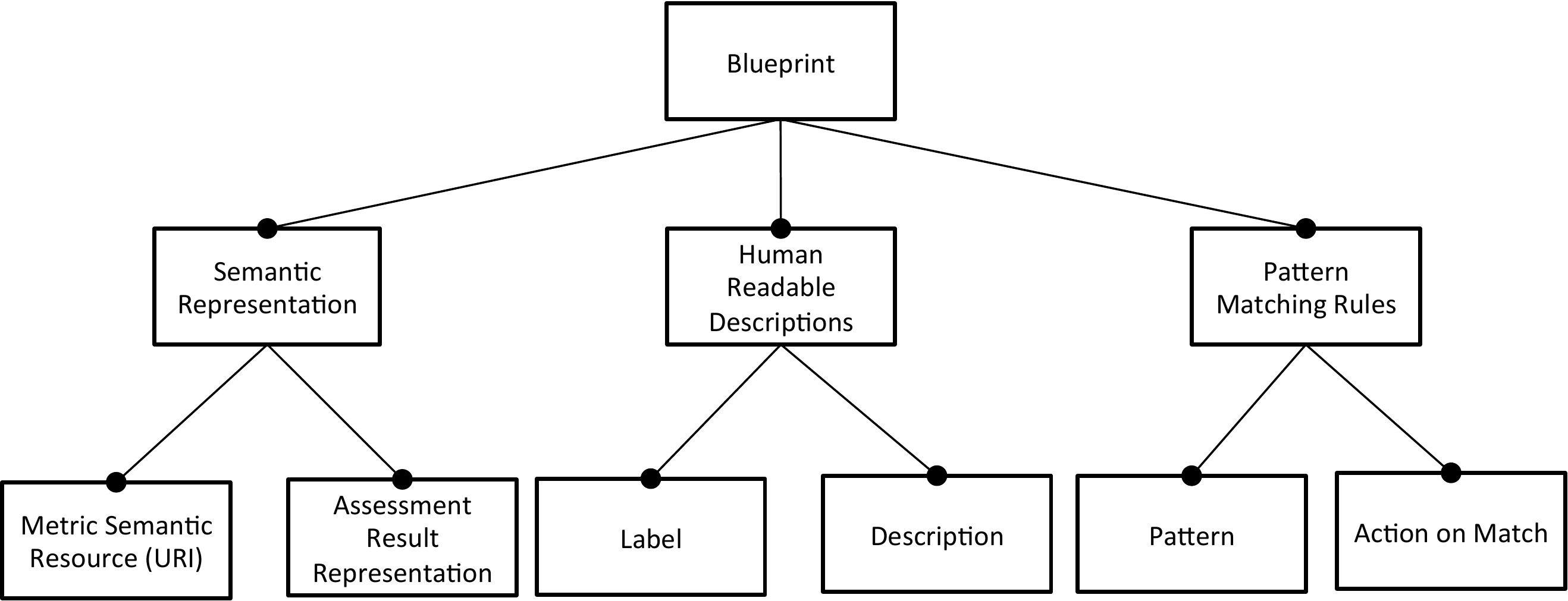} 
\caption{Feature Model for blueprints.} 
\label{fig:featureModel}
\end{figure}

\paragraph{Domain Terminology: }
A typical quality metric definition for linked open datasets consists of a \emph{pattern matching condition}, (i.e. matching the subject (\texttt{?s}), predicate (\texttt{?p}),  object (\texttt{?o}), or a mixture of these three with possibly advanced inspection), and an \emph{consequent action}.
This resembles the traditional \emph{if\dots then} statements of programming languages.
The full representation of an LQML metric definition is termed as \emph{blueprint}.
The feature model in Figure~\ref{fig:featureModel} describes the features required to create a quality metric blueprint.
A blueprint description should have enough information to assess a dataset based on the quality criteria (\emph{Pattern Matching Rules} in Figure~\ref{fig:featureModel}), and to enable the semantic description (\emph{Semantic Representation}) of the quality metadata for the criteria in question.
A description should also have a \emph{Human-Readable Description}.
This is required since blueprints will be shared amongst different data scientists and thus would enable anyone to understand complex patterns and actions.
All features are necessary in order to create one blueprint.

In a more formal manner, let ${r}$ be the root feature of a \emph{blueprint}.
$f_1$, $f_2$, $f_3$ represent the \emph{Semantic Representation} feature, \emph{Human Readable Description} feature, and \emph{Pattern Matching Rules} feature respectively such that:
\begin{equation}
r \longleftrightarrow f_1\land f_2\land f_3
\label{eq:formal_root}
\end{equation}
meaning that $f_1$, $f_2$, $f_3$ are mandatory features of $r$.

Similarly, 
\begin{equation}
f_1 \longleftrightarrow f_4\land f_5
\label{eq:formal_f1}
\end{equation}
where $f_4$ and $f_5$ represent the \emph{Metric Semantic Resource (URI)} and \emph{Assessment Result Respresentation} features respectively;
\begin{equation}
f_2 \longleftrightarrow f_6\land f_7
\label{eq:formal_f2}
\end{equation}
where $f_6$ represents the \emph{Label} feature and $f_7$ the \emph{Description} feature;
\begin{equation}
f_3 \longleftrightarrow f_8\land f_9
\label{eq:formal_f3}
\end{equation}
where $f_8$ represents the \emph{Pattern} feature and $f_9$ the \emph{Action on Match} feature.

\subsection{Design}
\label{sec:design}
Having identified the features for the proposed domain specific language, we here concisely describe its design and its features.
Mernik et al. \cite{Mernik:2005:DDL:1118890.1118892} describe a number of design patterns, three based on \emph{language exploitation} (designs based on existing languages), and another one for \emph{language invention}.
Our proposed DSL is based on the \emph{language invention} design pattern, where we fuse \todo{CL@JD: I don't understand what you mean by “a number of terms from the use cases available” – do you mean something like concrete URIs, representing resources from a use case specific dataset or terms from the use case specific domain? JD@CL: Better?}a number of specific terms (such as \emph{typeOf}) from the use cases available together with variable binding expressions used in the syntax of SPARQL  (i.e \texttt{?s ?p ?o}) to refer to specific elements in a triple.

\paragraph{Quality Metric (Blueprint) Structure: } 
A blueprint definition of a metric starts with the \texttt{def} keyword and has a rule semantics.
Each blueprint consists of the three features mentioned in Section~\ref{sec:analysis}.

\paragraph{Pattern Matching Rules Feature:}
Declarative patterns start with the keyword \texttt{match} (\emph{Pattern} feature).
If a triple matches the given condition, the given \texttt{action} (\emph{Action on Match} feature) is triggered.
Any input triple $(t_s,t_p,t_o)$ is matched against the conditions that follow the \texttt{match} keyword, enclosed into curly brackets (\{~\}).
A rule can have one or more conditions.
Conditions can be connected via the \emph{logical and} (\texttt{\&}) operator or the \emph{logical or} (\texttt{|}) operator.
Table~\ref{tbl:lhsConditions} shows possible conditions.

\begin{table}[h]
\centering
\resizebox{\textwidth}{!}{%
\begin{tabular}{ll}
\textbf{Name}                                  & \textbf{Description}                                                                                                                                                                                                                     \\ \hline
\multicolumn{1}{l|}{\texttt{typeof(?s  | ?o)}} & \begin{tabular}[c]{@{}l@{}}checks the type of subject or object. \\ \texttt{typeof(?s) == <U>} translates to the triple pattern ``?s a <U>''; \\ \texttt{typeof(?o) == <U> } translates to the triple pattern ``?o a <U>''.\end{tabular} \\ \hline
\multicolumn{1}{l|}{\texttt{?s | ?p == <U>}}   & matches the subject (\texttt{?s}) or the predicate (\texttt{?p}) against a given IRI (<U>).                                                                                                                                              \\ \hline
\multicolumn{1}{l|}{\texttt{?o == <U> | x}}          & matches the object (\texttt{?o}) against a given IRI (\texttt{<U>}) or a literal (\texttt{x}).                                                                                                                                                         
\end{tabular}
}
\caption{Pattern Matching Conditions}
\label{tbl:lhsConditions}
\end{table}

A condition can trigger one or more of the following \texttt{action}s:
\begin{itemize}
\item \texttt{map(?s,?o)} adds the subject and the object to a hash map as key/value (where the value is a list of objects);
\item \texttt{count} increments a counter;
\item \texttt{unique(?s | ?p | ?o)} increments a counter only if a unique instance of \texttt{?s}, \texttt{?p}, or \texttt{?o} is encountered.
\end{itemize}

In order not not limit LQML expressiveness, programmers can develop custom functions which can then be imported into Luzzu.
This enables data scientists to automatically use the imported functions in their defined \texttt{match} pattern.
Many domain specific languages, including XPath (cf. Section~\ref{sec:related}), not only have built-in functions\footnote{\url{http://www.w3.org/TR/xpath-30/\#id-function-calls}}, but also enable implementations to provide additional external functions.

\paragraph{Human Readable Descriptions:}
Descriptive human-readable comments are also required in these blueprints.
We provide the keywords \texttt{label} and \texttt{description} to provide the metric's name and its textual description; they translate to \texttt{rdfs:label} and \texttt{rdfs:comment}.

\paragraph{Semantic Representation:}
The definition also expects other information that describes a quality metric.
The \texttt{metric} (\emph{Metric Semantic Resource (URI)} feature) keyword expects a quality metric resource URI.
These resources are defined in a vocabulary that extends the Dataset Quality Ontology (daQ).
The \texttt{finally} (\emph{Assessment Result Representation} feature) keyword takes a defined function, and returns an output value, which is used as daQ observation value.

The \texttt{finally} keyword can have one of the following parameters:
\begin{itemize}
\item \texttt{actionresult(\textit{x)}} takes the value of the action, where \texttt{\textit{x}} stands for \texttt{map}, \texttt{count}, or \texttt{unique};
\item \texttt{ratio(x,y)} takes two parameters, which can either be integer or float numbers, or even a function (e.g. \texttt{count}) that returns a numeric value. The ratio function divides \texttt{x} by \texttt{y}.
\end{itemize}

\subsection{Implementation}
\label{sec:impl}

The LQML grammar is implemented in JavaCC (Java Compiler Compiler)\footnote{\url{https://java.net/projects/javacc}}.
JavaCC is a parser generator and a lexical analyser, where the grammar is specified in EBNF notation. 
Blueprints defined in LQML are interpreted by the JavaCC compiler where each blueprint is then interpreted and transformed into a Java class during Luzzu's runtime.

The following listing shows the EBNF grammar for the main parts of the LQML syntax.

\begin{lqml}[LQML EBNF grammar]{lst:bnf}
<Definition> ::= <Def> <Metric> <Label> <Description> <Match> <Action>
                 <Finally>

<Def> ::= "def" <LBrace> <Strict_Str> <RBrace> <Colon>

<Metric> ::= "metric" <LBrace> <IRIref> <RBrace> <SemiColon>

<Match> ::= "match" <LBrace> (<Condition>)+ <RBrace>

<Condition> ::=  <LParen> <TypeOf> | <DefinedFunction> | <other> <RParen>
                 (<logical_operator>)*

<TypeOf> ::= "typeof" <LParen> "?s" <RParen> <boolean_operator> <IRIref>

<other> ::= <LParen> "?s" <boolean_operator> <IRIref> <RParen> 
	| <LParen> "?p" <boolean_operator> <IRIref> <RParen>
	| <LParen> "?o" <boolean_operator> ( <IRIref> | <Quoted_Str> ) <RParen>  
	
<DefinedFunction> ::= <Strict_Str> "(" ("?s" | "?p" | "?o")* ")"

<IRIref> ::= refer to RFC 3987@~\cite{DueSui:iri05}@

<Action> ::= "action" <LBrace> ((<Map> | <Count> | <Unique>)(",")* )+
             <RBrace>

<Map> ::= "map"  <LParen> ("?s" | "?p" | "?o") ("?s" | "?p" | "?o")  <RParen>

<Count> ::= "count"

<Unique> ::= "unique" <LParen> ("?s" | "?p" | "?o")  <RParen>

<Finally> ::= "finally" <LBrace> (<Number>  | <ActionResult> <Ratio>)+
              <RBrace>

<ActionResult> ::= "actionresult" <LParen> ("map" | "count" | "unique")
                   <RParen> 

<Ratio> ::= "ratio" <LParen> (<Number> | <NumericFunction>) "," ( <Number> | <NumericFunction>) <RParen>

<NumericFunction> ::= <Map> | <Count> | <Unique>
\end{lqml}

\paragraph{External Functions:}
External functions, defined as Java classes, are preloaded into Luzzu beforehand.
These can only be used within a \texttt{match} pattern.
The structure (as described in the EBNF \texttt{<DefinedFunction>}) requires a function name (as a string) and zero or more variables.

\subsection{Blueprint Examples}
In order to keep up the quality within the EFO, domain experts from the EBI (cf. Section~\ref{sec:analysis}) defined relevant quality metrics.
One relevant metric is that they identify a percentage of how many resources are actually defined as sub-classes (\texttt{rdfs:subClassOf}) of other classes.
Listing~\ref{lst:structure} shows an LQML metric definition for the above.

\begin{lqml}[EBI Use Case Example in LQML]{lst:structure}
def{SubClassCounter}:
  metric{<http://www.example.org/ebiqm#SubClassCountingMetric>};
  label{"SubClassCountingMetric"};
  description{"Provides a measure for counting the number of resources that are defined as sub-classes"};
  match{(?p == <http://www.w3.org/2000/01/rdf-schema#subClassOf>)};
  action{count, unique(?s)};
  finally{ratio(actionresult(count), actionresult(unique))}.
\end{lqml}

One of Data Publica's requirements is that each resource they define has a human readable description or label.
This means that they quantify a percentage of how many resources have either an \texttt{rdfs:label} or an \texttt{rdfs:comment} defined.
This metric is defined by LQML in Listing~\ref{lst:structure1}.

\begin{lqml}[Data Publica Use Case Example in LQML]{lst:structure1}
def{HumanReadableLabel}:
metric{<http://www.example.org/dpqm#SubClassCountingMetric>}; 
label{?Human Readable Labelling Metric"};
description{"Provides a measure for identifying the @ratio@ of human readable labels of defined resources in a dataset?};
match{(typeof(?s) ==  <http://www.example.org/dp#Class>) && ((?p == <http://www.w3.org/2000/01/rdf-schema#@label@>) || (?p == <http://www.w3.org/2000/01/rdf-schema#comment>)))}; 
action{count, unique(?s)};
finally{ratio(actionresult(count), actionresult(unique))} .
\end{lqml}




\section{Sharing Blueprints – Luzzu Blueprint Ontology}
\label{sec:extending}

We envisage that blueprint descriptions can be stored and shared in a common pool of metrics, similar as how different users of the IFTTT service\footnote{\emph{If This Than That} is an online service allowing users to create simple rules that trigger events: \url{https://ifttt.com/}} can share rule recipes.
Shared blueprints can be either reused or modified to fit the purpose of another use case.
The LQML language itself does not enable such sharing to be done with ease.
For this purpose, we propose an ontology, the Luzzu Blueprint Ontology (prefix: \texttt{lbo}), which facilitates the semantic representation of LQML blueprints.
Exploiting this option, the ontology enables us to distribute blueprints as semantic resources.
Moreover, these semantic resources are easily queried and visualised.

\begin{figure}[tb]
\center
\includegraphics[width=\columnwidth]{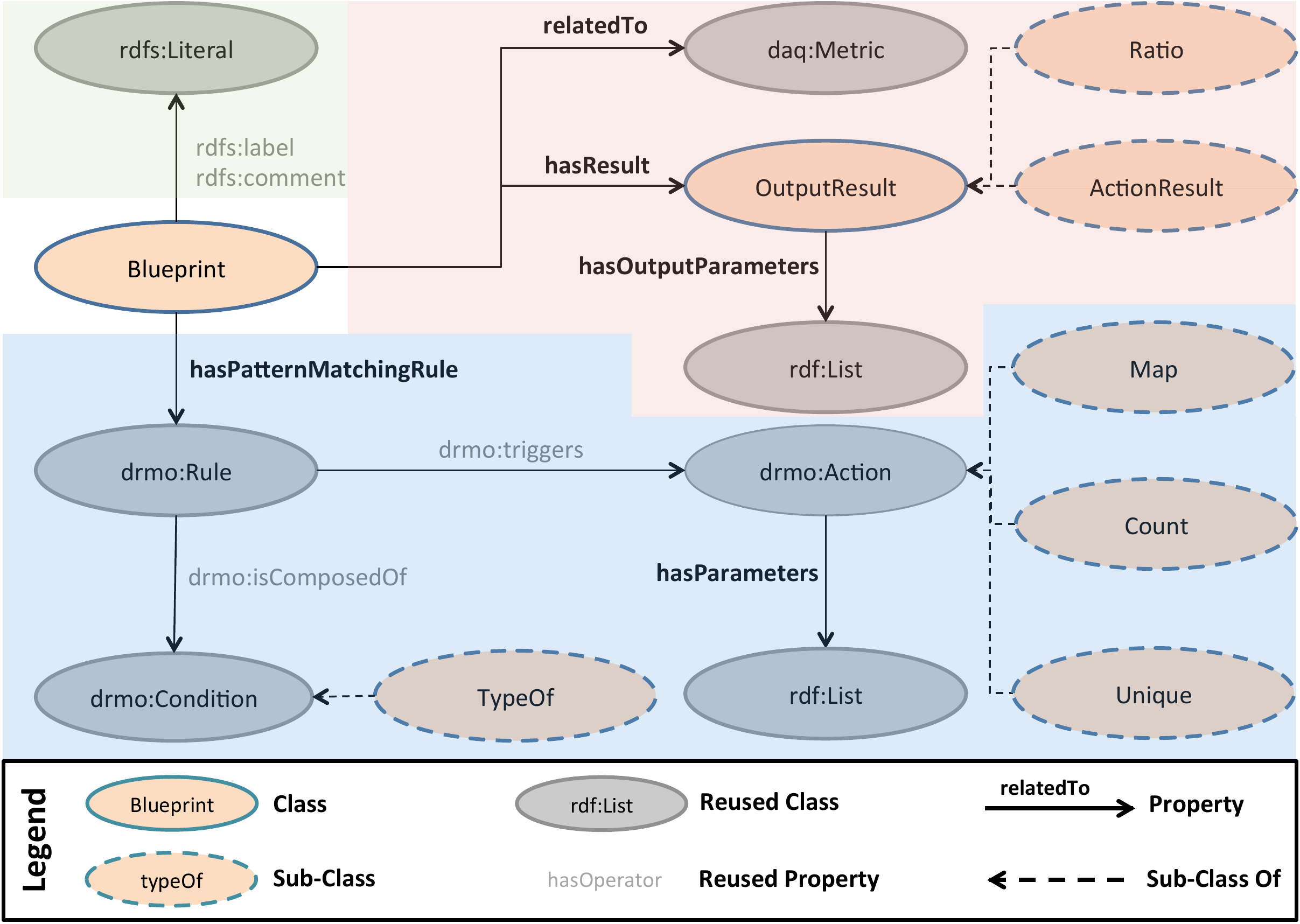} 
\caption{Luzzu Blueprint Ontology.} 
\label{fig:lbo}
\end{figure}

In line with the Semantic Web principles, the Luzzu Blueprint Ontology (depicted in Figure~\ref{fig:lbo}\todo{CL@JD: In my PDF viewer (SumatraPDF on Windows; Sören also uses it) the stroke of the left arrow ``subclass of'' is invisible.}) reuses known concepts and domain specific ontologies.

\todo{CL@JD: This listing is \emph{so} trivial that I'd skip this, and rather have this triple as the first triple of a longer listing (to which we can refer here).}Listing~\ref{lst:root} shows an RDF definition of the \textit{root} feature (cf. Figure~\ref{fig:featureModel}), Blueprint (defined as \texttt{lbo:Blueprint} in the proposed ontology).
\begin{lstlisting}[language=N3,caption={Defining the root feature $r$ in Turtle notation},label={lst:root}]
lbo:Blueprint a rdfs:Class .
\end{lstlisting}

The vocabulary incorporates the three main \textit{features} described in Section~\ref{sec:analysis}; \textit{Semantic Representation} (highlighted in light red -- top right), \textit{Human Readable Descriptions} (highlighted in light green -- top left), and \textit{Pattern Matching Rules} (highlighted in light blue -- bottom part).
Therefore, the formal definitions described in Equations~\ref{eq:formal_root},~\ref{eq:formal_f1},~\ref{eq:formal_f2}, and ~\ref{eq:formal_f3} are given a semantic RDF definition with the Luzzu Blueprint Ontology.

For the \textit{human readable descriptions}, we make use of the standard \texttt{rdfs:label} and \texttt{rdfs:comment} properties to represent the blueprint's label and description respectively.
From the W3C RDF Schema definition of \texttt{rdfs:label} and \texttt{rdfs:comment}, these two properties are used to provide a human-readable name and description of a resource respectively. 
Both properties must have an \texttt{rdfs:Resource} as a domain (which instances of \texttt{lbo:Blueprint} automatically are) and a literal (\texttt{rdfs:Literal}) as its range, i.e. a textual value.
These two properties facilitate the Semantic Web notation for Equation~\ref{eq:formal_f2}.

The \textit{semantic representation} (cf. Listing~\ref{lst:formal_f1}) of a quality metric is represented in an instance of the blueprint ontology using the proposed properties \texttt{lbo:relatedTo} and \texttt{lbo:hasResult}.
The former links a blueprint instance to a \texttt{daq:Metric} resource.
The latter represents the resulting (\texttt{finally} in terms of LQML) output.
For this purpose we also introduced two sub-classes of the concept \texttt{lbo:OutputResult}; \texttt{lbo:Ratio} and \texttt{lbo:ActionResults}.
The semantics for the latter sub-classes were discussed in Section~\ref{sec:design}.
An \texttt{lbo:OutputResult} can have one or more output parameters.
The proposed property \texttt{lbo:hasOutputParameters} expects a resource of type \texttt{rdf:List} as its range.
\begin{lstlisting}[mathescape,language=N3,caption={Defining the \emph{Semantic Representation} feature, Equation~\ref{eq:formal_f1}, using a Semantic Web notation},label={lst:formal_f1}]
# Representing the Metric Semantic Resource (URI) feature $f_4$
lbo:relatedTo a rdf:Property ;
	rdfs:domain lbo:Blueprint ;
	rdfs:range daq:Metric .
	
# Representing the Assessment Result Representation feature $f_5$
lbo:hasResult a rdf:Property ;
	rdfs:domain lbo:Blueprint ;
	rdfs:range lbo:OutputResult .
	
# ... definitions of OutputResult, subclasses and hasOutputParameters property.
\end{lstlisting}

The Digital.me Rule Management ontology (DRMO) enables users to express rules in terms of resources and concepts that are available in a personal knowledge base~\cite{Debattista2013Processing}.
Although it is expected to operate in a closed-world environment, its flexibility of being a domain-independent ontology enables its reuse in other scenarios.
The DRMO concepts are inspired by the Event-Condition-Action (ECA) pattern, where the latter pattern is used in event-driven architectures.
In light of the \emph{pattern matching rules} feature, we make use of the DRMO concepts, creating a sub-class of \texttt{drmo:Condition}, \texttt{lbo:TypeOf}; and three sub-classes of \texttt{drmo:Action}, namely \texttt{lbo:Map}, \texttt{lbo:Count}, and \texttt{lbo:Unique}.
The representation of a condition (the \textit{Pattern} feature -- $f_8$), is left intact.
On the other hand, we extended the concept \texttt{drmo:Action} the \emph{Action on Match} feature - $f_9$) with a property named \texttt{lbo:hasParameters}, where the range of the newly proposed property is a resource of type \texttt{rdf:List}.
The property \texttt{hasPatternMatchingRule} (cf. Listing~\ref{lst:formal_f3}) defines the described extended \texttt{drmo:Rule} of an \texttt{lbo:Blueprint}.
The semantics of the \texttt{drmo:Rule} concept, together with the properties \texttt{drmo:isComposedOf} and \texttt{drmo:triggers}, gives a semantic definition for Equation~\ref{eq:formal_f3}.
\begin{lstlisting}[mathescape,language=N3,caption={Defining the \emph{Pattern Matching Rules} feature $f_3$},label={lst:formal_f3}]
# Representing the relationship between the root feature ($r$) and the Pattern Matching Rules feature $f_3$
lbo:hasPatternMatchingRule a rdf:Property ;
	rdfs:domain lbo:Blueprint ;
	rdfs:range drmo:Rule .
	
# ... definitions of subclasses for Action (Map, Count, Unique), and hasParameters extension to DRMO.
\end{lstlisting}

Debattista et al.~\cite{Debattista2013Processing} provide a mechanism that transforms \texttt{drmo:Rule}s into SPARQL queries. 
Therefore, blueprint patterns can be transformed into SPARQL queries and reused even within quality frameworks, such as RDFUnit~\cite{kontokostasDatabugger}, employing a SPARQL engine as their main assessment tool.
This makes Luzzu blueprints interoperable with other quality assessment frameworks.



\section{Evaluation}
\label{sec:evaluation}
In order to assess the usability of LQML, we gauge the language systematically against the ``cognitive dimensions of notation'' (CD) evaluation framework, a methodology developed in~\cite{Blackwell2001}
This evaluation framework has previously been applied to Semantic Web languages (e.g. \cite{LPHTM:SemanticWebPipes09}).
These dimensions provide a comprehensive view of how users can manage and use a defined language.
Each dimension describes a specific aspect in relation to the language notation.
Blackwell and Green~\cite{Blackwell02notationalsystems} describe the following 13 dimensions:

\textbf{Viscosity} questions the effort required by the user to lead out a change.\\
\emph{Assessment:} LQML metrics can be defined using a simple text editor.
Each statement is defined for a particular definition (blueprint) and is not related to other definitions. 
Therefore, changing a statement in a definition does not require a change in any other place, thus resulting in a low viscosity.

\textbf{Premature Commitment} measures any planning required before leading out a task.\\
\emph{Assessment:} Based on declarative programming, LQML users only require to define rules based on the patterns they want to match. 
Also, declarations are not required before a blueprint definition. 
The only premature commitment is that metrics have to be defined in an ontology (whose URI is defined in the blueprint definition) based on the daQ ontology.

\textbf{Hidden Dependencies} measures if dependencies are specifically indicated in all existing directions.\\
\emph{Assessment:} Blueprint definitions cannot be connected to each other, therefore each definition has a fixed rule and action, together with other descriptions.

\textbf{Error-proneness} measures the possibility of users making mistakes while using the language.\\
\emph{Assessment:} A definition is made up of only six components. 
This means the learning curve is not too steep.
However, since these six components must always be fixed in the same order, i.e. \texttt{metric}, \texttt{label}, \texttt{description}, \texttt{match}, \texttt{action}, \texttt{finally}, there is an increased possibility of the user making a mistake, but this is mitigated by error messages from the LQML parser.

\textbf{Abstraction} measures high level concepts which are not easily grasped by the users, since they do not refer to concrete instances. 
This dimension thus measures the language's abstraction level.\\
\emph{Assessment:} In LQML, we try to keep the number of keywords at the pattern matching feature to a minimum, such that users can have full control on their declarative patterns.
In this way there is a very low level of abstraction. 

\textbf{Secondary Notation} indicates the availability of options for encoding extra context information within the syntax itself, such as comments. \\
\emph{Assessment:} A definition requires a description; further important information can be added in an unstructured way as comments (starting with \texttt{\#}, extending to the end of the line).

\textbf{Closeness of Mapping} measures the degree of similarity between the representation language and the real-world domain.\\
\emph{Assessment:} Our aim is to try to simplify the definition of metrics as much as possible, keeping in mind that possible non-Java experts are using this tool.
Despite having this beneficial feature that widens the tool's audience, expert users who require to create more complex metrics, for example, calculating the response time of a server serving a resource, must implement LQML extension functions in Java; metrics with complex matching conditions and actions will even have to be implemented completely in Java.

\textbf{Consistency} measures the usability of the language; in other words, how easy is it for a user to write similar blueprints once the notation pattern has been learned. \\
\emph{Assessment:} Unlike in the error-proneness dimension, we here consider that the fixed syntax structure is actually a feature, in a way that consistency is kept for all blueprint definitions. 

\textbf{Diffuseness} measures the space required by the notation; i.e. the amount of workspace occupied by the language.\\
\emph{Assessment: } Although the blueprints themselves have a clear goal, the rules within the definition might be messy and unclear since different conditions have to be defined in brackets. 
\todo{CL@JD: If LQML doesn't even give \texttt{\&} precedence over \texttt{|}, is this a bug (given that all programming languages do it), or is it a feature (as we explicitly target non-programmers, who might not be aware of this anyway)? JD@CL: not sure what you mean by this}In LQML, users have to define the precedence of evaluating the conditions (using brackets). 
The fact that LQML blueprints are defined in a simple text editor means that users might find some difficulty in understanding a rule.

\textbf{Progressive Evaluation} measures the understandability of the language even for a solution that is incomplete. The possibility to try out a partial solution helps users in further understanding their work\\
\emph{Assessment: } It is possible to incrementally refine definitions by, e.g., starting with a partial match and a simple “count” action, and then to further refine the matching pattern by adding conditions, and to define a more complex action.

\textbf{Role Expressiveness} indicates the language's notation and its expressiveness vis-a-vis the whole solution.\\
\emph{Assessment: } Our tool is aimed towards the definition of quality metrics for linked data. In a definition, all required information is adequately labelled to enable easy identification. 

\textbf{Visibility} measures the degree of visibility of the language's notation. 
If concepts are encapsulated into concepts of a more abstract level, this reduces the visibility of the notation.\\
\emph{Assessment: } All available notation is directly visible to the user. 


\textbf{Provisionality} measures the ability of the language to allow users to explore potential options.\\
\emph{Assessment: }
Similarly to the secondary notation dimension, potential options can be explored by temporarily commenting out parts of a definition.

Together the assessment w.r.t.\ these dimensions provides a comprehensive heuristic guide of LQML, particularly focusing on language features that have not been implemented in an immediate response to the given quality assessment requirements.
From this evaluation we can identify certain problems in the current implementation of the syntax, such as the possibility of reusing components of blueprints within others. 
These heuristics also stress the importance of the need of a better presentation view tool (graphical interface) for the user, while also highlighting that whilst we are widening the scope of metric definition for non-Java experts, we are limiting ourselves to simple pattern matching metrics and thus more complex metrics cannot be defined. 
These measurements will help us in the second phase of the language definition.



\section{State of the Art}
\label{sec:related}

A Domain Specific Language (DSL) is a small declarative programming language focusing on a particular domain, offering appropriate notations and abstractions, in a way that is easy to use for non-programmers~\cite{vanDeursen:2000:DLA:352029.352035,dsl-little-languages,Mernik:2005:DDL:1118890.1118892}.
The authors of~\cite{vanDeursen:2000:DLA:352029.352035,dsl-little-languages} describe a number of benefits of DSLs, including:
\begin{itemize}
\item the enhancement of productivity; 
\item the incorporation of domain knowledge;
\item the possibility of portability;
\item the understandability of declarative programs by domain experts themselves;
\item easily maintainable code.
\end{itemize}
A DSL development methodology starts with the \emph{decision} stage, where stakeholders decide if the effort in investing in a new DSL pays off in the future.
If the stakeholders decide to go ahead, then they proceed to the \emph{analysis} stage, where the problem domain is identified and knowledge about that domain is gathered.
Following that, the DSL is then \emph{design}ed where the knowledge is concisely described as semantic notations and graphically by using tools such as a feature model\footnote{\url{http://en.wikipedia.org/wiki/Feature_model}}.
Finally, the DSL is \emph{implement}ed.
In the article ``When and How to Develop Domain-Specific Languages'', Mernik et al.~\cite{Mernik:2005:DDL:1118890.1118892} provide the reader with a comprehensive insight on DSL development methodologies, by identifying patterns for the four stages of the development methodology.

In this section we mention a few examples from a growing list of domain specific languages.
Each DSL builds on a data model to encapsulate domain knowledge into an abstract notation.

Domain specific languages are popular within various applications. 
\LaTeX~is a document preparation typesetting system usually used for technical and scientific publications.
HTML and XML are generic markup languages that are also DSLs.
The former is used to generate websites, whilst the latter is used as an interoperable data model.
XPath\footnote{\url{http://www.w3.org/TR/xpath-30/}} is an expression language enabling the processing of values in an XML data model.
XPath uses path expressions to navigate through XML.
RuleML\footnote{\url{http://ruleml.org}} is an XML markup language that allows rules to be defined using a formal notation.

In the Semantic Web there are a number of domain specific languages.
The RDF data model\footnote{\url{http://www.w3.org/TR/rdf11-primer/}} is based on triple statements (subject–predicate–object) that enable the description of real-world objects as machine-readable semantic resources.
On top of this data model, RDF Schema\footnote{\url{http://www.w3.org/TR/rdf-schema/}} is a vocabulary that provides a number of classes and properties to describe a resource in RDF.
This schema language also provides the basic concepts for the development of new domain specific ontologies.
The RDF data model can be serialised in different formats, such as RDF/XML and Turtle.
SPARQL\footnote{\url{http://www.w3.org/TR/rdf-sparql-query/}} is a domain specific query language for querying the RDF data model.
The Web Ontology Language (OWL)\footnote{\url{http://www.w3.org/TR/owl2-overview/}} adds more semantics on top of the RDF Schema.
OWL enables users to create inferencing rules and statements on RDF.
Going further away from the data model, the Rule Interchange Format (RIF)~\cite{conf/rr/Kifer08} is a web standard defining an interchange language for rules within different systems to achieve interoperability.
It focuses on the definition of various dialects, which enables the exchange of rules within different systems.
The above mentioned DSLs are just a few examples tackling different aspects of the RDF data model.
Similar to these, the proposed Luzzu Quality Metric Language is also based on this semantic data model.


\section{Concluding Remarks}
\label{sec:conclusion} 

Data quality assessment is crucial for the wider deployment and use of Linked Data.
With the Luzzu Quality Metric Language we empower domain experts who are not proficient in using third generation programming languages to define domain specific quality metrics for linked open datasets.
We defined the Luzzu Blueprint Ontology to ensure that quality metrics defined with our proposed domain specific language can be shared, queried and reused easily in a semantic manner.
LQML was evaluated systematically against the Cognitive Dimensions of Notation, a methodology developed purposely to assess formal notations such as those of programming languages.
The evaluation pointed out shortcomings in the current implementation of the DSL and possible future improvements.

Together with the Luzzu framework, we see this as the first step to break the poor quality reputation barrier of Linked Open Datasets.
Regarding future work, we aim to create an interactive user interface for the definition of LQML metrics in order to visualise and author metrics, and to offer an online pool of quality metrics that can be queried and downloaded into Luzzu, and to which the Linked Open Data Community can contribute.
We also plan to refine LQML with more generic keywords that can be used within the \texttt{match}, \texttt{action}, and \texttt{finally} parts of a definition.




\bibliographystyle{splncs03}
\bibliography{paper,../../bib/eis,../../bib/aksw,../../bib/external/kwarc} 

\end{document}